# Zero-Shot Prognostic Covariate Adjustment and Mechanistic Week-1 Myelosuppression Updating Across 9,500 Oncology Patients Under FDA Guidance

Varshal Patel[1,2], Neema Darabi[1,2]

[1]Carle Illinois College of Medicine, University of Illinois Urbana-Champaign, Urbana, IL, USA; [2]Renmara

Correspondence: Varshal Patel, varshal2@illinois.edu

The FDA's May 2023 guidance on covariate adjustment recommends "that sponsors adjust for covariates that are anticipated to be most strongly associated with the outcome of interest" and notes that "it may be useful to use previous studies (e.g., a Phase 2 trial) to select prognostic covariates or form prognostic indices." Its August 2024 guidance on dosage optimization, part of Project Optimus, states that "the dose- and exposure-response relationships should be evaluated to select a dosage(s) for clinical trial(s)." We analyzed the control arms of 14 randomized trials from Project Data Sphere, with 5,074 patients with overall survival follow-up, in non-small cell lung cancer (NSCLC), pancreatic ductal adenocarcinoma (PDAC) and metastatic colorectal cancer (mCRC), and in three AACR Project GENIE Biopharma Collaborative (BPC) cohorts (4,440 patients). In all, about 9,500 patients were studied. We used the regularized baseline prognostic index to compute out-of-fold estimates of the share of covariate-adjusted log-rank variance removed ($\rho^2$), and found these to be 0.177 to 0.186 for the three indications. For the locked held-out trials, the value was 0.183 (second-line NSCLC docetaxel), 0.177 (first-line PDAC gemcitabine) and 0.110 (first-line mCRC FOLFOX plus bevacizumab). Refitting in GENIE BPC resulted in estimates of 0.073 to 0.127 for first-line and 0.122 to 0.219 for later lines. In a simulated 600-patient trial, 5,000 within-stratum re-randomizations gave a type I error rate of 0.051 for the adjusted test. At $\rho^2 = 0.18$, adjustment raises power from 80.0% to 87.2%, cuts required events by 18% or shrinks the control arm by about 30%. A Friberg marrow prior updated with week 1 blood counts raised the AUROC for new cycle 1 grade 3 or worse neutropenia from 0.791 to 0.816 ($n = 3{,}465$; calibration slope 0.999; Brier skill 0.190) and raised it in 5 of 8 cross-trial rows (up to +0.116).

## 1 Introduction and regulatory framework

Late-stage oncology trials may use chemotherapy control arms with patients who have wide spreads of tumor burden, organ function and inflammatory state. Consequently, the standard error of the log-rank statistic is inflated. For example, a trial might be powered for a hazard ratio of 0.75, yet the observed hazard ratio is 0.85. In this situation, the primary end point might not be met even if the drug works. To address this problem, investigators could adjust for a prognostic index trained on baseline data from the control arms of earlier late-stage trials. As an example, in Project Data Sphere, a clinical trials data repository, we identified 14 randomized trials and used their control arms. The data included 5,074 patients with overall survival follow-up. In combination regimens, neutropenia and thrombocytopenia drive cycle 1 dose holds, dose reductions, and discontinuations. In our data, the day 1 baseline covariates explain at most about 20% of within-trial variance in continuous cycle 1 nadir depth. A forecast of each patient's cycle 1 toxicity risk, sharpened within the first week of treatment, could serve both trial monitoring and dosage justification. In August 2024 the U.S. Food and Drug Administration (FDA) issued final guidance on dosage optimization for patients with oncologic diseases, part of its Project Optimus [1]. It observes that dose-finding trials designed to determine the maximum tolerated dose "may not adequately consider other data, such as low-grade symptomatic toxicities (i.e., grade 1-2), dosage modifications, drug activity, pharmacokinetics, pharmacodynamics, and dose- and exposure-response relationships."

The FDA's final guidance of May 2023, "Adjusting for Covariates in Randomized Clinical Trials for Drugs and Biological Products" [2], says that "it may be useful to use previous studies (e.g., a Phase 2 trial) to select prognostic covariates or form prognostic indices," and on timing, that "sponsors should prospectively specify the detailed procedures for executing covariate-adjusted analysis before any unblinding of comparative data. FDA review will emphasize the prespecified primary analysis rather than post-hoc analyses using different models or covariates." A single pre-specified index built on external trials can therefore enter the primary analysis as one covariate [3].

The present paper has four main results. In Section 2, we analyze published estimators and their operating characteristics. In Section 3, we introduce and evaluate a cross-trial prognostic index. In Section 4, we introduce and validate a Friberg-type marrow model for predicting toxicity from week 1 blood counts. In Section 5, we discuss several limitations of the public data used in this paper and outline a deployment design.

## 2 Estimators for survival covariate adjustment

Let $n$ patients be randomized to experimental ($A_i = 1$) or control ($A_i = 0$) treatment with probability $\pi$ within strata $S_i \in \{1,\dots,K\}$. Denote by $X_i$ the baseline data, $N_i(t)$ the counting process and $Y_i(t)$ the at-risk process for patient $i$. The prognostic index $m(X_i)$ is a regularized cross-trial baseline prognostic index mapping pre-randomization (Day $\leq$ 1) CDISC ADaM subject characteristics (ADSL), routine screening laboratory panels (ADLB: hematologic, hepatic, renal, and inflammatory/tumor markers), and baseline RECIST target-lesion burden (ADTR), locked via SHA-256 prior to comparative unblinding. The index enters every estimator below as one scalar covariate, $\hat{m}_i = m(X_i)$. In the words of the FDA guidance, baseline covariates "refer to demographic factors, disease characteristics, or other information collected from participants before the time of randomization" [2].

### *2.1 Covariate-adjusted stratified log-rank test*

The stratified log-rank numerator is

$$U = \sum_{s=1}^{K} \sum_{i \in s} \int_0^\infty \left\{ A_i - \frac{\bar{Y}_{1s}(t)}{\bar{Y}_s(t)} \right\} dN_i(t), \tag{1}$$

where $Y_{1s}$ and $Y_s$ count experimental and all patients at risk in stratum $s$. Under $H_0$, $U = \sum_s \sum_{i \in s} (A_i - \pi) M_i + o_p(\sqrt{n})$, with $M_i = N_i(\infty) - \int_0^\infty Y_i(t)\, d\Lambda_s(t)$ the martingale residual from the pooled within-stratum Nelson-Aalen estimator [4]. Ye, Shao and Yi [5] remove from this score its projection on baseline covariates. With the single covariate $\hat{m_i}$,

$$U_{\text{adj}} = U - \sum_s \sum_{i \in s} (A_i - \pi)\, \tilde{\beta}_s\, (\hat{m}_i - \bar{m}_s), \qquad \tilde{\beta}_s = (1-\pi)\hat{\beta}_{1s} + \pi\hat{\beta}_{0s}, \tag{2}$$

where $\hat{\beta}_{as}$ is the least-squares slope of $M_i$ on $\hat{m_i}$ among arm-$a$ patients in stratum $s$ and $\bar{m_s}$ is the stratum mean of the index. The statistic $Z_{\text{adj}} = U_{\text{adj}}/\hat{\sigma}_{\text{adj}}$ uses their randomization-based variance estimator, which under $H_0$ and simple randomization within strata is approximately

$$\hat{\sigma}^2_{\text{adj}} = \sum_s \sum_{i \in s} \pi(1-\pi)\big\{\hat{M}_i - \tilde{\beta}_s(\hat{m}_i - \bar{m}_s)\big\}^2. \tag{3}$$

$Z_{\text{adj}}$ is asymptotically standard normal under $H_0$ for any working slope, which is the formal sense in which misspecification cannot inflate type I error [6, 7], and its asymptotic variance never exceeds that of the unadjusted test [5]. Under $H_0$ the two arm slopes coincide and

$$\frac{\operatorname{Var}(U_{\text{adj}})}{\operatorname{Var}(U)} \to 1 - \rho^2, \qquad \rho^2 = \operatorname{Corr}\{M, \hat{m}(X) \mid S\}^2, \tag{4}$$

the squared within-stratum correlation between the martingale residual and the index. We estimate $\rho^2$ out of fold on control arms, where it is a property of the index and the population rather than of any treatment. The guidance warns that “an analysis ignoring stratified randomization is likely to overestimate standard errors and can be unduly conservative when performing inference for the average treatment effect,” and that “covariate adjustment can still be performed with covariates that are not prognostic, but there may not be any gain in precision (or may be a loss in precision) compared with an unadjusted analysis” [2]. That loss applies to some estimators and to finite samples; asymptotically, the adjusted log-rank statistic cannot lose precision.

### 2.2 ANCOVA for restricted mean survival time

For the restricted mean survival time $\text{RMST}(\tau) = E\{\min(T,\tau)\} = \int_0^\tau S(t)\, dt$ we use Lin’s interacted ANCOVA [8] on jackknife pseudo-observations $Y_i$ of $\min(T_i,\tau)$ [9],

$$Y_i = \alpha + \theta A_i + \gamma(\hat{m}_i - \bar{m}) + \eta A_i(\hat{m}_i - \bar{m}) + \varepsilon_i, \tag{5}$$

with stratum indicators added and Huber-White sandwich standard errors. Centering the index and including the interaction keep $\theta$ consistent for the marginal RMST difference without assuming the linear model, and

$$\operatorname{Var}(\hat{\theta}_{\text{adj}}) \approx (1-\rho^2)\operatorname{Var}(\hat{\theta}_{\text{unadj}}), \tag{6}$$

with $\rho^2 = \operatorname{Corr}\{\min(T,\tau), \hat{m}(X)\}^2$ for this estimator. This matches the guidance, which states that “even when the linear regression model is misspecified and does not accurately capture the relationships between the outcome, covariates, and treatment, covariate adjustment through a linear model is a valid method for estimating and performing inference for the average treatment effect” [2]. With the interaction included, the Huber-White standard errors remain valid or conservative [8]. It also notes that “in trials with time-to-event outcomes, the hazard ratio can also be non-collapsible,” one reason to report the RMST difference alongside the log-rank test [2].

### 2.3 Operational translations at $\rho^2 = 0.18$

*Power at fixed sample size and events.* A test designed for power $1 - \beta$ under the unadjusted variance attains

$$1 - \beta_{\text{adj}} = \Phi\Big(\frac{z_{1-\alpha/2} + z_{1-\beta}}{\sqrt{1-\rho^2}} - z_{1-\alpha/2}\Big). \tag{7}$$

The FDA guidance expects that adjusting for baseline covariates “will generally reduce the variability of estimation of treatment effects and thus lead to narrower confidence intervals and more powerful hypothesis testing” [2]. At $\rho^2 = 0.18$, 80.0% power becomes 87.2% and 90.0% becomes 94.7%. We next simulated a first-line pancreatic ductal adenocarcinoma (PDAC) trial with 600 patients, true conditional HR 0.72 and stratified into the Eastern Cooperative Oncology Group (ECOG) performance status of 0 vs 1 or more and liver metastases. The index was locked and applied unchanged, and the within-stratum $\rho^2$ was 0.196. With the unadjusted stratified log-rank test, the observed HR was 0.87, with a z value of −1.55 and p value of 0.121 (missed). The adjusted test on the same data gave $z = -2.00$ and $p = 0.045$, with a variance ratio of 0.81, and Lin’s RMST(12) estimator had a standard-error ratio of 0.89.

*Required events and readout time.* Schoenfeld’s formula [10] gives $D = 4(z_{1-\alpha/2} + z_{1-\beta})^2/(\log\text{HR})^2$ and adjustment scales it as

$$D_{\text{adj}} = \lceil D(1-\rho^2) \rceil. \tag{8}$$

With a two-sided alpha of 0.05 and 90% power, the number of deaths needed for HR = 0.72 is 390, or 320 at $\rho^2 = 0.18$, 18% fewer. We assumed 600 patients over a 18-month accrual period, with a 9-month median survival in the control arm. The event-driven analysis then moves from month 26.0 to month 21.5, and across $\rho^2 = 0.177$ to 0.186 the registrational readout comes 4.4 to 4.6 months earlier. The guidance leaves the choice open: “the sample size and power calculations can be based on adjusted or unadjusted methods. The latter will often lead to a more conservative sample size” [2].

*Control-arm size.* Holding the experimental arm at $n$ patients under 1:1 allocation and requiring the adjusted variance to match the unadjusted 1:1 design gives

$$n'_c = n\,\frac{1-\rho^2}{1+\rho^2}, \tag{9}$$

We estimated that the control arm would be 30.5% smaller at $\rho^2 = 0.18$. Applied to the three held-out trials, Equation (9) gives a control arm 19.9% to 30.9% smaller (see Figure 1). The assumption is that the variance explained in both arms is similar, which is untestable with the control-arm data alone. So this is a design option built on $\rho^2$, not a validated result. The three translations are shown in Figure 1 over the whole range of $\rho^2$ that we observed in Section 3.

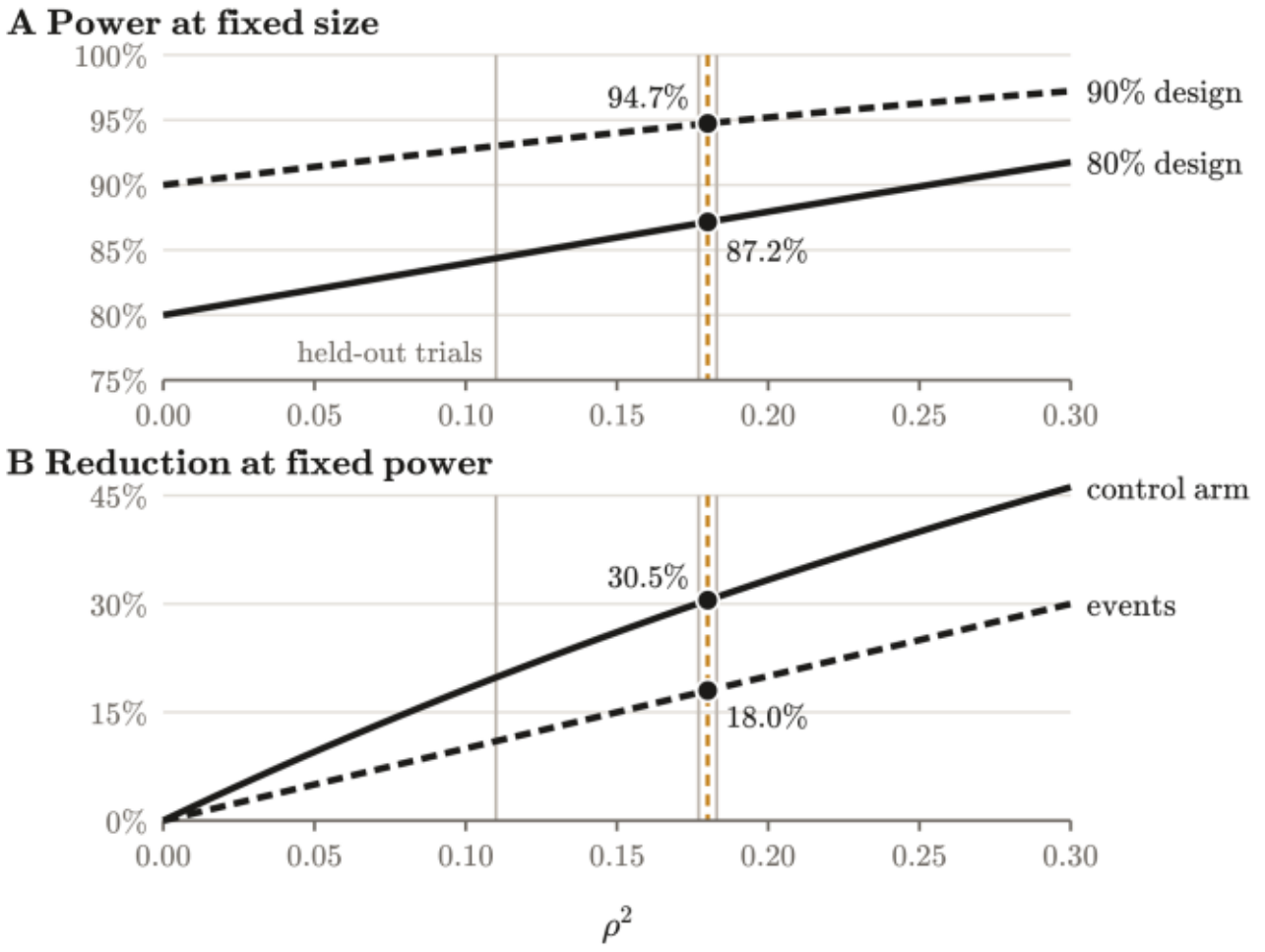

**Figure 1.** Operating characteristics as a function of $\hat{\rho}^2$, from Equations (7) to (9). A: power of trials designed for 80% and 90% power under the unadjusted variance. B: percent fewer required events (dashed) and percent smaller control arm with the experimental arm fixed (solid). Grey lines: the three held-out trials ($\hat{\rho}^2 = 0.110, 0.177, 0.183$). Dashed amber line: $\hat{\rho}^2 = 0.18$.

## 3 Empirical validation across 9,500 oncology patients

### 3.1 Data and validation tiers

All randomized trials were identified from Project Data Sphere [11], an open-access platform which collects, curates, and shares de-identified patient-level data from historical oncology trials. We used the control arms of 14 of its randomized trials: 4 trials in the non-small cell lung cancer (NSCLC), 5 trials in the PDAC, and 5 trials in the metastatic colorectal cancer (mCRC) patient cohorts. Carboplatin plus paclitaxel (first line), squamous gemcitabine plus cisplatin (first line), erlotinib (previously treated), and docetaxel, VITAL [12] (second line) were the NSCLC control arms. PDAC trials include 4 trials of gemcitabine arms (first line, among them MPACT and VANILLA) and 1 trial of capecitabine or 5-FU (second line). The 5 trials of mCRC included 4 trials of FOLFOX (first line, 2 with bevacizumab) and 1 trial of FOLFIRI (second line). A total of 12 Phase 3 trials, 1 Phase 2 trial (PDAC), and 1 Phase 1/2 trial were included in the analysis. Of the 5,074 overall survival (OS) analyzable patients, 1,536 patients had NSCLC, 1,151 patients had PDAC, and 2,387 patients had mCRC.

For the evaluation of the model, three validation tiers were defined. Tier 1 is the out-of-fold cross-validation, in which an indication-pooled model was trained on all of that indication's control arms, and each patient was scored by a model blind to that patient. Tier 2 validation is leave-one-trial-out (LOTO), where a model is trained on the other development trials of the same indication, and the held-out trial is scored, no refit of the model takes place. The model was tested on locked held-out trials (Tier 3), VITAL (PDS133), LEAP (PDS186) and HORIZON III (PDS78). Tier 3 validation used 3 held-out trials that were excluded from feature selection, tuning and thresholds once the split was declared (Section 5.3). The lock comprises the index serialized and SHA-256 hashed; the hash is checked at scoring. Scorings were written to an append-only ledger with one scoring per trial. Uncertainty bands are 80% intervals from a 500-draw patient bootstrap of the out-of-fold predictions.

**Table 1.** Overall-survival variance removed ($\hat{\rho}^2$) by the baseline prognostic index, by validation tier. Base: age, sex, ECOG, stage (GENIE: age, sex, stage, prior lines, regimen class). $\Delta n_c$: control-arm reduction from Equation (9).

| *Cohort* | *n* | *Deaths* | *Base* | *Index (80% band)* | $\Delta n_c$ |
|---|---|---|---|---|---|
| *Out-of-fold development cross-validation, pooled* | | | | | |
| NSCLC, 4 trials | 1,536 | 1,172 | 0.048 | 0.177 (0.154, 0.199) | 30.1% |
| PDAC, 5 trials | 1,151 | 918 | 0.041 | 0.186 (0.164, 0.211) | 31.4% |
| mCRC, 5 trials | 2,387 | 1,358 | 0.080 | 0.180 (0.163, 0.197) | 30.5% |
| *Leave-one-trial-out transport, 9 development trials* | | | | | |
| Range | | | | 0.065 to 0.198 | |
| *Cryptographically locked held-out test trials (zero-shot)* | | | | | |
| PDS133 (2L NSCLC) | n.r. | 346 | 0.072 | 0.183 (0.149, 0.222) | 30.9% |
| PDS186 (1L PDAC) | n.r. | 158 | 0.044 | 0.177 (0.119, 0.235) | 30.1% |
| PDS78 (1L mCRC) | 690 | 241 | 0.031 | 0.110 (0.080, 0.141) | 19.9% |
| *AACR GENIE BPC, refit out of fold* | | | | | |
| NSCLC, 1L | 518 | 363 | 0.024 | 0.082 (0.052, 0.111) | 15.1% |
| CRC, 1L | 858 | 486 | 0.057 | 0.127 (0.102, 0.154) | 22.6% |
| PDAC, 1L | 662 | 575 | 0.019 | 0.073 (0.046, 0.102) | 13.6% |
| NSCLC, 2L+ | n.r. | 229 | 0.050 | 0.122 (0.079, 0.176) | 21.8% |
| CRC, 2L+ | 703 | 443 | 0.030 | 0.218 (0.183, 0.253) | 35.8% |
| PDAC, 2L+ | 510 | 446 | 0.076 | 0.219 (0.170, 0.264) | 36.0% |

Held-out regimens: PDS133 docetaxel; PDS186 gemcitabine; PDS78 FOLFOX plus bevacizumab. Pooled rows include each indication's held-out trial, scored out of fold; stratified by trial, the pooled values are 0.169 (NSCLC), 0.178 (PDAC) and 0.135 (mCRC). n.r.: not reportable under small-cell suppression.

### 3.2 Variance removed

In the out-of-fold test, the models achieved a $\hat{\rho}^2$ value of 0.177 in NSCLC, 0.186 in PDAC, and 0.180 in mCRC (Table 1, Figure 2). Because pooled cohorts mix trials with different median OS, they were also stratified by trial: 0.169 for NSCLC, 0.178 for PDAC, and 0.135 for mCRC. When the LOTO is applied to the data from the 9 randomized development trials with OS follow-up, the $\hat{\rho}^2$ was 0.065 to 0.198 (Table 1, Figure 2).

In the held-out control arms, 346 patients died in the VITAL trial, 158 in the LEAP trial, and 241 out of 690 patients in the HORIZON III trial. On these held-out trials, the locked index removed 0.183 in VITAL, 0.177 in LEAP and 0.110 in HORIZON III (Table 1). In the VITAL trial, the share of log-rank variance removed was 7.2% after including age, sex, performance status and stage. The full locked index, which adds the screening labs and the baseline tumor burden, removed 18.3%. A model tuned and cross-validated in VITAL itself reached 0.179, about the same as the cross-trial index (0.183) without in-study training. No development trials that used docetaxel were involved during the training of the model. The FDA guidance notes that "randomization is often stratified by baseline covariates" and that "a covariate adjustment model should generally include strata variables and can also include covariates not used for stratifying randomization" [2]. Through Equation (9), these held-out values correspond to control arms 30.9%, 30.1% and 19.9% smaller.

We also used the AACR Project GENIE Biopharma Collaborative, whose releases cover 1,109 patients with PDAC (v1.0), 1,485 with CRC (v2.0), and 1,846 with NSCLC (v2.0) [13, 14]. All patients had undergone clinical sequencing at one of four cancer centers (Memorial Sloan Kettering, Dana-Farber, Vanderbilt-Ingram, Princess Margaret). Control episodes were standard-of-care regimens started in metastatic disease, one per patient per setting, with delayed entry for sequencing after treatment start. The registry data did not include ECOG and screening lab results. Therefore, we trained the model without the ECOG and trial screening labs, using treatment history, metastatic sites, histology, and genomic profile out of fold within GENIE, so the index was not transported. The GENIE $\hat{\rho}^2$ for the first-line cohorts ranged from 0.073 to 0.127, and for the later-line cohorts ranged from 0.122 to 0.219. Later-line cohorts included patients with immune checkpoint inhibitor therapy in an earlier line. The gain over design variables was highest in the later lines of treatment (0.030 to 0.218 in CRC), where treatment history and the pace of earlier progression carry much of the prognosis.

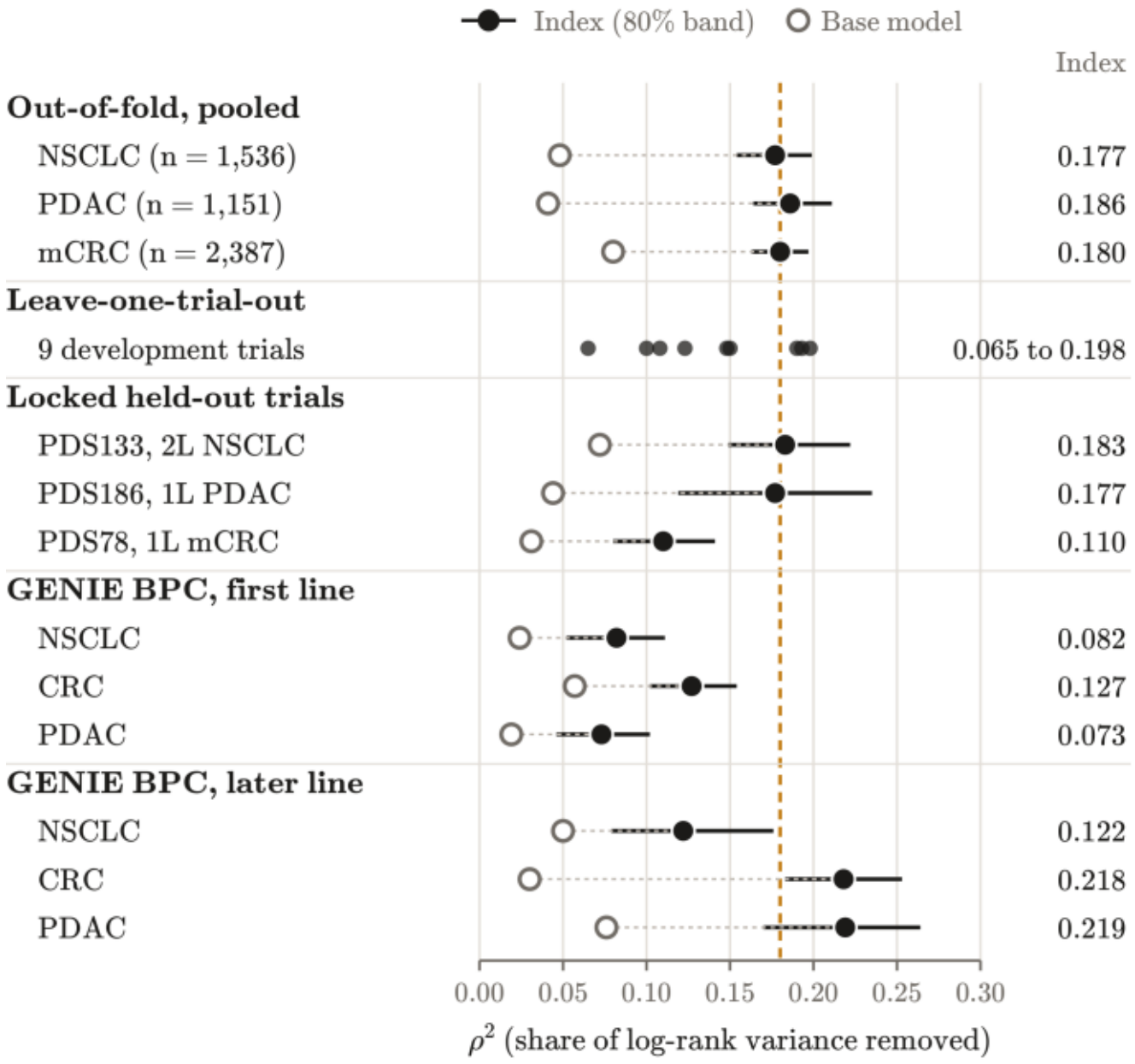

**Figure 2.** Overall-survival variance removed ($\rho^2$) by validation tier, with values of the index on the right. Filled points: baseline prognostic index with 80% patient-bootstrap bands. Open points: base model (age, sex, ECOG, stage; GENIE: age, sex, stage, prior lines, regimen class). The LOTO row shows the nine held-out development trials. Dashed line: $\rho^2 = 0.18$.

Type I error checking was performed by re-randomizing the simulated trial of Section 2.3. Specifically, we randomly permuted the labels within each stratum (with 4 strata) 5,000 times and reran both tests. The adjusted test rejection rate was 0.0512 (95% CI 0.045 to 0.057), which was close to the nominal level of 0.050, and the unadjusted was 0.052. The p-value of the Kolmogorov-Smirnov test was 0.90, suggesting that the adjusted p-values are uniformly distributed, consistent with nominal type I error for the simulated trial. The FDA guidance notes that "sponsors can conduct randomization/permutation tests with covariate adjustment" [2]; here re-randomization served only to check the type I error of the model-based tests.

## 4 Mechanistic Week-1 myelosuppression updating

### *4.1 Model*

For each lineage (absolute neutrophil count, ANC, and platelets) the prior is a 5-compartment hematopoietic and epithelial transit ODE prior (following Friberg et al. [15]) inverted via low-dimensional numerical quadrature conditioned on routine Week-1 (Day $\leq$ 8) complete blood counts and conformalized via Jackknife+ [16]. The marrow chain is a proliferating pool $P$, three transit compartments and circulating cells $C$:

$$\begin{aligned} \frac{dP}{dt} &= k_{\mathrm{prol}} P\{1 - E(t)\}\Big(\frac{\mathrm{Circ}_0}{C}\Big)^{\gamma} - k_{\mathrm{tr}} P, \\ \frac{dT_1}{dt} &= k_{\mathrm{tr}}(P - T_1), \quad \frac{dT_2}{dt} = k_{\mathrm{tr}}(T_1 - T_2), \\ \frac{dT_3}{dt} &= k_{\mathrm{tr}}(T_2 - T_3), \quad \frac{dC}{dt} = k_{\mathrm{tr}} T_3 - k_{\mathrm{circ}} C, \end{aligned} \tag{10}$$

with $k_{\mathrm{tr}} = 4/\mathrm{MTT}$, $k_{\mathrm{prol}} = k_{\mathrm{tr}}$ at homeostasis, each patient starting from its own baseline count $\mathrm{Circ}_0$, and $E(t)$ a monotone drug effect driven by the Cycle-1 schedule. Day-1 covariates set prior weights $\pi_j(X_i)$ on quadrature nodes $\theta_j$ for the patient's sensitivity and kinetics. Counts $y_{ik}$ drawn on Days $t_{ik} \leq 8$ update the weights by their likelihood,

$$w_{ij} \propto \pi_j(X_i) \prod_k \phi\Big(\frac{\log y_{ik} - \log C(t_{ik}; \theta_j)}{\sigma}\Big), \tag{11}$$

and the risk of new-onset severe cytopenia after the Day-8 landmark is $\hat{p}_i = \sum_j w_{ij} \Pr\{\min_{9 \leq t \leq 42} C(t; \theta_j) < c^*\}$, with $c^*$ the Common Terminology Criteria for Adverse Events (CTCAE) threshold. Every tuning and calibration step was refit inside each training fold.

### *4.2 Cohort, outcomes and out-of-fold performance*

The cohort consisted of control-arm patients with a recorded cycle 1 dose. Patients were at risk of a new severe cytopenia on days 9 to 42, counted from a landmark at day 8 among patients whose baseline was at or above the threshold and who had no value below it on days 2 to 8. The thresholds were ANC below $1.0 \times 10^9$/L (grade 3 or worse), ANC below $0.5 \times 10^9$/L (grade 4), and platelets below $50 \times 10^9$/L (grade 3 or worse). The week 1 posterior (S4) adds to the day 1 baseline model (S1) the counts drawn on days 2 to 8 and the week 1 adverse events with their onset dates. A week 1 draw was available for 45% of at-risk patients.

**Table 2.** Out-of-fold Week-1 updating in the development trials: new-onset severe cytopenia on Days 9 to 42. S1: Day-1 baseline model. S4: Week-1 posterior. BSS: Brier skill score.

| *Outcome* | *Events* | *AUROC S1* | *AUROC S4* | *BSS S1* | *BSS S4* | *Slope* |
|---|---|---|---|---|---|---|
| ANC $< 1.0$, G$\geq 3$ | 454 | 0.791 | **0.816** | 0.159 | 0.190 | 0.999 |
| ANC $< 0.5$, G4 | 123 | 0.793 | **0.813** | 0.081 | 0.085 | 0.921 |
| PLT $< 50$, G$\geq 3$ | 43 | 0.675 | **0.735** | 0.006 | 0.020 | 0.835 |

ANC rows: 3,465 at-risk patients; platelet row: 3,558. Thresholds in $10^9$/L. PLT: platelets. Slope: calibration slope of S4.

Out of fold, the grade 3 or worse neutropenia prediction models, which were based on the information of the week 1 update, showed an area under the receiver operating characteristic curve (AUROC) of 0.816, up from 0.791 at day 1, a Brier skill of 0.190, up from 0.159, and a calibration slope of 0.999 (Table 2). For grade 4 neutropenia, the AUROC went from 0.793 to 0.813, whereas for grade 3 or worse thrombocytopenia, it went from 0.675 to 0.735. Moreover, without week 1 adverse events, the AUROC of grade 3 or worse neutropenia prediction was 0.815, so the blood counts carry nearly all of the gain.

### *4.3 Zero-shot transport and the unseen-regimen guardrail*

Cross-trial AUROC values were calculated for events vs. non-events for the eight cross-trial rows with at least 11 events and 11 non-events (Table 3). The AUROC of PDS132 for first-line PDAC platelets rose from 0.606 to 0.722 (a gain of 0.116; 12 events) and that of PDS108 for first-line NSCLC neutropenia rose from 0.649 to 0.718; these were two of the five rows in which the AUROC went up. The update also repaired calibration in held-out PDS133, where day 1 risks transported badly: the Brier skill scores went from −4.14 to −0.16 for grade 3 or worse neutropenia. The Brier skill score for PDS133 for grade 4 neutropenia went from −3.46 to 0.04. Three rows declined. PDS263, a FOLFIRI trial with no week 1 draws on record, had a lower AUROC for grade 3 or worse neutropenia after the update. PDS108 had a lower AUROC for grade 4 neutropenia, and PDS272 had a lower AUROC for grade 4 neutropenia (57 and 11 events).

**Table 3.** Zero-shot transport of Week-1 updating: every cross-trial row with at least 11 events and 11 non-events. S1, S4: AUROC of the Day-1 model and of the Week-1 posterior.

| *Trial* | *Outcome* | *Events* | *Wk-1 %* | *S1* | *S4* | Δ | *BSS S1, S4* |
|---|---|---|---|---|---|---|---|
| PDS132 | PLT $< 50$ | 12 | 83.7 | 0.606 | 0.722 | +0.116 | 0.03, 0.04 |
| PDS108 | ANC $< 1.0$ | 173 | 86.7 | 0.649 | 0.718 | +0.069 | 0.09, −0.01 |
| PDS133† | ANC $< 0.5$ | 31 | 12.4 | 0.627 | 0.677 | +0.050 | −3.46, 0.04 |
| PDS133† | ANC $< 1.0$ | 48 | 12.4 | 0.635 | 0.674 | +0.038 | −4.14, −0.16 |
| PDS263 | ANC $< 0.5$ | 15 | 0.0 | 0.637 | 0.641 | +0.004 | −0.24, −0.005 |
| PDS263 | ANC $< 1.0$ | 50 | 0.0 | 0.644 | 0.632 | −0.012 | −1.35, 0.01 |
| PDS108 | ANC $< 0.5$ | 57 | 86.7 | 0.728 | 0.696 | −0.032 | 0.01, 0.007 |
| PDS272 | ANC $< 0.5$ | 11 | 92.5 | 0.664 | 0.615 | −0.049 | −0.34, −0.06 |

†Locked held-out test trial; other rows are LOTO, trained on all other trials. Δ: S4 minus S1, computed before rounding. Wk-1 %: share of at-risk patients with a count on Days 2 to 8. PDS132: 1L PDAC gemcitabine; PDS108: 1L NSCLC carboplatin plus paclitaxel; PDS133: 2L NSCLC docetaxel; PDS263: 2L mCRC FOLFIRI; PDS272: 1L squamous NSCLC gemcitabine plus cisplatin.

The unseen regimen test used the PDS133 control arm, with single-agent docetaxel, which was absent from all development control arms. The day 1 penalized ridge regression, trained on all development trials, had an R squared of −1.319 (Spearman 0.16) for the continuous cycle 1 log-nadir depth in PDS133, far worse than predicting the mean. The same day 1 regression fitted within single trials explained at most 0.198 of this variance. The mechanistic prior alone had an R squared of −0.060 (Spearman 0.10). However, combined with the baseline covariates and read at the realized dosing and draw days, the R squared was 0.291 (Spearman 0.66), and 0.164 with planned dosing (Figure 3). The exposure ridge, given the same realized dosing, had R squared −1.559. Because the realized-day readings use recorded dosing and sampling dates, they measure the value of the mechanistic prior rather than a day 1 forecast; the day 1 comparison shows the failure the prior prevents. Project Optimus describes dose selection based on "a preliminary understanding of dose- and exposure-response (such as for safety, tolerability, and activity)" as a more informed approach [1].

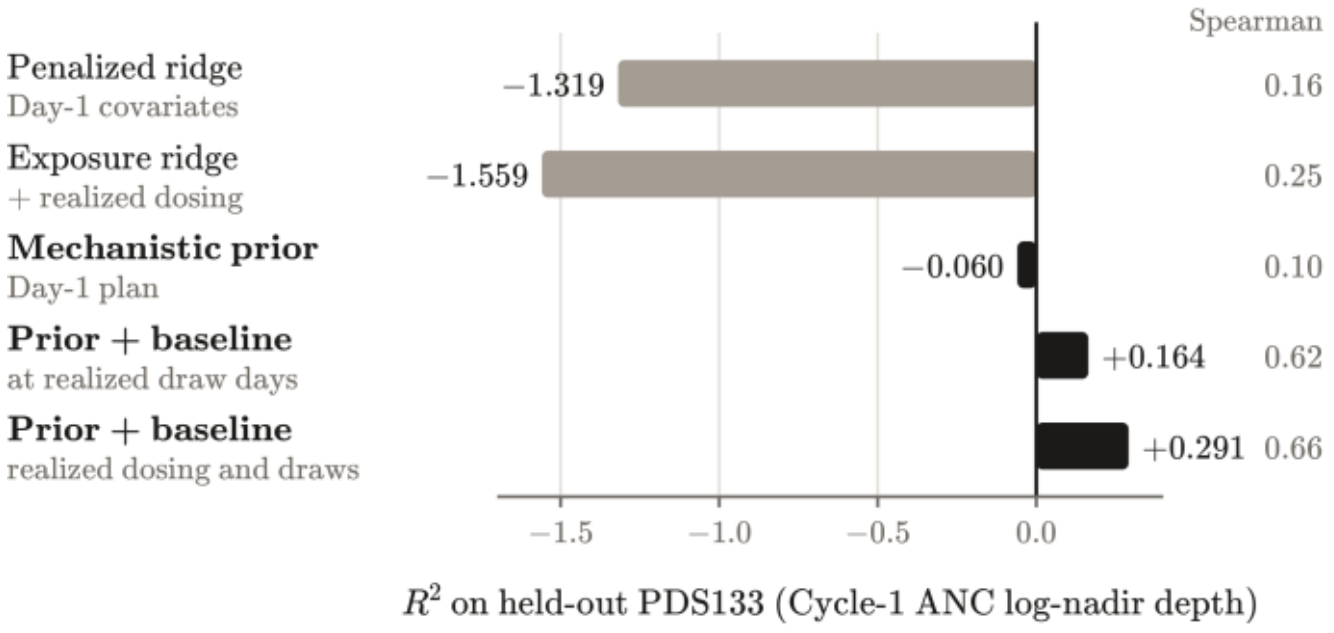

**Figure 3.** Unseen-regimen guardrail on held-out PDS133 (single-agent docetaxel, absent from every development control arm): $R^2$ for continuous Cycle-1 ANC log-nadir depth, all models trained on the development trials. Grey: regression baselines. Black: mechanistic prior. The two lower bars read the prior at the recorded blood-draw days and are ablations, not Day-1 forecasts.

## 5 Multi-cycle confounding boundary and zero-egress deployment

### *5.1 Dynamic dose-modification policies*

Project Optimus asks that "safety and tolerability should be compared across the multiple dosages, including: duration of exposure; proportion of patients who are able to receive all planned doses; percentage of patients that require dosage interruptions, dose reductions, and drug discontinuations for adverse reactions; time to the first dosage modification…" [1]. About 2,650 patients were evaluated, and 8 randomized development trials (3 mCRC, 1 NSCLC, 4 PDAC) were included. Dose modifications in cycles 2 to 6 were assessed with 5 sequential decisions to delay, reduce, or withhold or omit subsequent cycles. The modification was defined as delay of 7 days or more, dose below 85% of starting dose, or hold or omission. Following Kennedy [17], the incremental propensity score (IPS) intervention multiplies the odds of modification by $\delta$ at each decision, $q_\delta(g) = \delta g/\{1 + (\delta - 1)g\}$. Continuation, death, and withdrawal were left to follow their natural course.

The target, 12-month mortality under the shifted policy, was estimated by cross-fitted doubly robust sequential regression with inverse-probability-of-censoring weights. The bias against exact discrete-tree synthetic oracles was found to be below $10^{-12}$. It also has a 93% to 99% coverage over 100 simulated cohorts of the real size. The real data showed that the variance was 1.8 to 2.3 times lower than inverse probability weighting.

The toxicity-guided policy, which shifts the odds of modification only after CTCAE grade 2 to 3 marrow or organ toxicity in the completed cycle, showed no detectable effect on 12-month mortality: +0.32 percentage points at $\delta = 2.00$ (95% CI −0.41 to +1.05). When decisions about therapy changes were shifted at every decision, 12-month mortality was higher by 1.69 percentage points (95% CI +0.47 to +2.90). However, the association may have been due to confounding by indication (doses held in declining patients), and adding restaging scans, tumor markers and organ chemistry did not remove it. Separating drug exposure from tumor progression needs multi-dose plasma pharmacokinetics (ADaM ADPC) and serial circulating tumor DNA, which public control arms lack and sponsor studies hold. This is also why Project Optimus prefers randomized dose comparisons: "Randomization (rather than enrolling patients to non-randomized dosage cohorts) promotes comparability of patients receiving each dosage, minimizing bias in estimation of dose- and exposure-response relationships" [1].

### *5.2 Zero-egress deployment*

Each analysis is packaged as a deterministic container that runs inside the sponsor's virtual private cloud. The container starts with `--network none`. The sponsor's CDISC ADaM tables (ADSL, ADTTE, ADLB, ADEX, ADAE, ADTR) are mounted read-only (they do not change) and only the aggregate outputs are written. Before comparative unblinding, the container emits a lock certificate with the SHA-256 digests of each input table and the serialized scorer. At final analysis, the statistical analysis plan (SAP) covariate can be checked byte for byte against the lock certificate. In testing, synthetic ADaM data was run end to end, with zero network calls. A sponsor can run a retrospective evaluation on a completed study and lock the SAP covariates for the next one, so patient-level data never leaves the sponsor's environment (Figure 4).

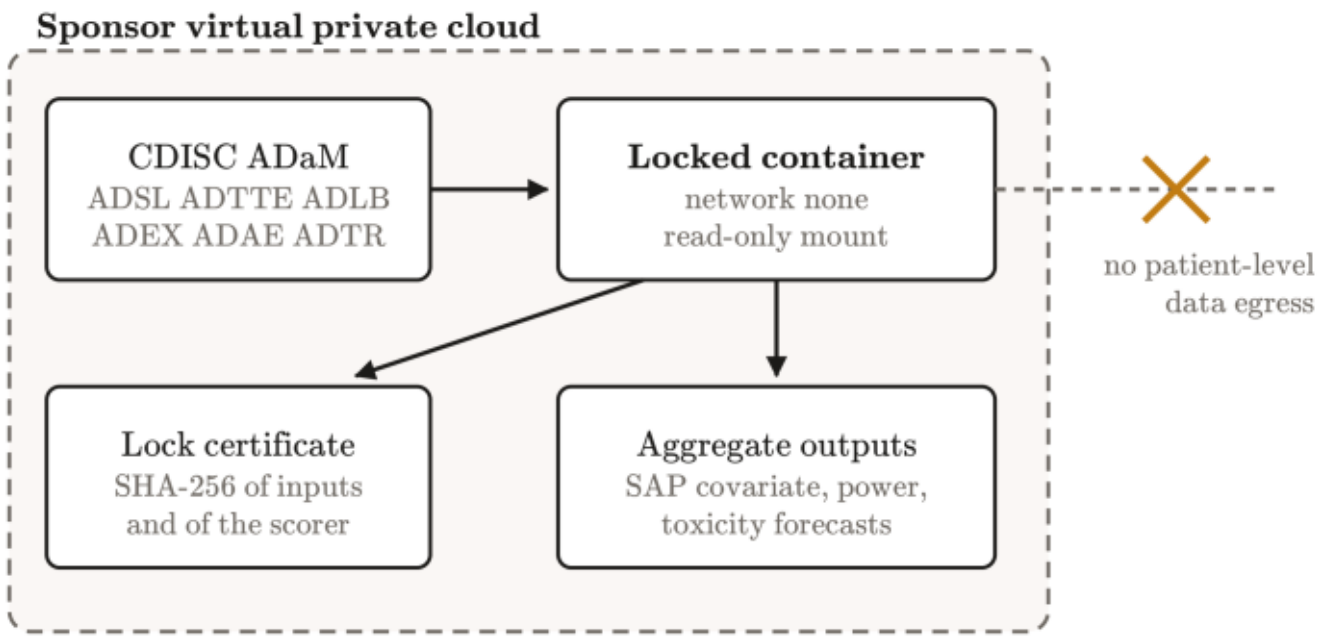

**Figure 4.** Zero-egress deployment. The container runs inside the sponsor's virtual private cloud with networking disabled and a read-only ADaM mount; only aggregate outputs and the SHA-256 lock certificate leave it.

### *5.3 Limitations*

The Project Data Sphere validation used control arms only, so $\rho^2$ provides a measure of prognostic strength, not treatment effect heterogeneity. The feature definitions and tuning grids were fixed before the held-out split and all trials were available, but the held-out trials were never used for selection afterwards, and each was scored once. GENIE BPC results come from observational regimen cohorts and a refit index. Two-arm and type I error results are based on simulations with the specific assumptions made. Week 1 updating requires that scheduled early draws be performed, and calibration after transport remained imperfect. The dose-modification estimates assume no unmeasured confounding given the recorded history, an assumption the unguided result suggests is not met.

## 6 Conclusion

A single pre-specified prognostic index from routine baseline data, locked before unblinding, removed about 18% of the log-rank variance on the held-out randomized control arms. Under the 2023 guidance, the index is worth seven points of power, 18% fewer events, or a control arm about 30% smaller, with no cost in type I error. A mechanistic marrow prior, updated on week 1 counts, sharpened cycle 1 toxicity forecasts and kept predictions near the mean on an unseen regimen. Both analyses can run inside a sponsor's own environment. As the 2023 guidance puts it, "covariate adjustment leads to efficiency gains when the covariates are prognostic for the outcome of interest in the trial" [2].

### Data and acknowledgments

This publication is based on research using information obtained from www.projectdatasphere.org, which is maintained by Project Data Sphere. Neither Project Data Sphere nor the owner(s) of any information from the web site have contributed to, approved or are in any way responsible for the contents of this publication. The authors acknowledge the American Association for Cancer Research and its financial and material support in the development of the AACR Project GENIE registry, as well as members of the consortium for their commitment to data sharing; interpretations are the responsibility of the authors. Only aggregate statistics are reported, and cells of 1 to 10 patients are suppressed (n.r.).

### Competing interests

The authors are co-founders of Renmara, which develops the methods described.